\documentclass[preprint, 12pt, amsmath, amssymb, aps, superscriptaddress, showkeys]{revtex4-2}
\usepackage[utf8]{inputenc}
\usepackage[colorlinks=true,linkcolor=blue,citecolor=blue,urlcolor=blue]{hyperref}
\usepackage{slashed}
\usepackage{color}
\usepackage{graphicx}
\usepackage{multirow}
\usepackage{booktabs}
\graphicspath{{pics/}}
\allowdisplaybreaks[1]

\begin{document}
\title{Unraveling alignment pattern in high-energy particles via transverse momentum disbalance analysis}

\author{Igor Lokhtin}
\email{lokhtin@www-hep.sinp.msu.ru}
\affiliation{Skobeltsyn Institute of Nuclear Physics, Lomonosov Moscow State University, Moscow, Russia}
\author{Aleksei Nikolskii}
\email{alexn@theor.jinr.ru}
\affiliation{Skobeltsyn Institute of Nuclear Physics, Lomonosov Moscow State University, Moscow, Russia}
\affiliation{Bogoliubov Laboratory of Theoretical Physics, Joint Institute for Nuclear Research, Dubna, Russia}
\author{Alexander Snigirev}
\email{snigirevam@my.msu.ru}
\affiliation{Skobeltsyn Institute of Nuclear Physics, Lomonosov Moscow State University, Moscow, Russia}
\affiliation{Bogoliubov Laboratory of Theoretical Physics, Joint Institute for Nuclear Research, Dubna, Russia}

\begin{abstract}
The hypothesis of the relation between strong azimuthal correlations, known as alignment, in  families of hadrons and photons observed in cosmic ray experiments and the selection procedure of the highest-energy particles together with the transverse momentum conservation are tested in the framework of the HYDJET++ model. The results show that the high degree of alignment can appear in nucleus-nucleus interactions at reasonable values of the transverse momentum disbalance of most energetic detected particles.
\end{abstract}

\keywords{Properties of Hadrons; Specific QCD Phenomenology; Relativistic Heavy Ion Physics; Particle Correlations and Fluctuations; Quark-Gluon Plasma; Cosmic Rays}

\maketitle

\section{Introduction}
\label{sec:intro}

In our previous study~\cite{Lokhtin:2023tze,LokhtinYF24}, within the framework of the geometrical approach, we demonstrated that a high degree of alignment can partly arise  due to the selection procedure for the highest-energy particles, the energy deposition threshold, and the conservation of transverse momentum. This promising result encourages us to explore this hypothesis further within the framework of a realistic model of hadron interactions. The phenomenon in question involves the observation of strong angular correlations in families of hadrons and photons, which are products of interactions between cosmic rays and target nuclei~\cite{pamir1,pamir2,pamir3,pamir4,book}. This phenomenon could potentially indicate a coplanar structure of the observed events and was termed alignment~\cite{pamir4}. Among various angular characteristics, alignment most clearly demonstrates the position of the most energetic registered particles (or their clusters) relative to a straight line in the plane of the emulsion film. Additionally, strong angular correlations of particles have been observed in stratospheric~\cite{strat1,strat2} X-ray-emulsion chamber experiments. To date, there is no universally accepted explanation for this intriguing phenomenon, despite numerous attempts to find one (see, for instance,~\cite{book,halzen,man,mukh,Lokhtin:2005bb,Lokhtin:2006xv,deroeck} and references therein).

The long-range azimuthal correlations, also known as the ridge effect, observed in heavy ion collisions at RHIC~\cite{phobos} and in high-multiplicity $pp$ interactions at the LHC~\cite{cms}, stimulated the search for any manifestations~\cite{Dremin:2010yd,Lokhtin,mukh2} of the alignment phenomenon under LHC conditions. It is presumed that alignment should already appear at the LHC, as the total collision energy of the beams in these experiments exceeds the energy threshold $\sqrt{s_{\rm eff}} \geq 4$ TeV for detected particles exhibiting alignment, as seen in experiments with cosmic rays.
However, the determination of any correspondence between these striking phenomena was unsuccessful because they were observed in quite different rapidity intervals, practically without overlapping, and in different reference frames. Moreover, subsequent studies have provided an explanation for the ridge effect within established theoretical approaches: for instance, as a simple interplay between elliptic and triangular flows~\cite{Eyyubova:2014dha}. This fact is largely consistent with a rather widespread point of view (questioned in~\cite{mukh,mukh3,Mukhamedshin:2019dus}) that the alignment phenomenon is nothing more than a tail in the distribution caused by fluctuations.
In this paper, we present a development and continuation of the ideas described in our recent studies~\cite{Lokhtin:2023tze, LokhtinYF24}, but our approach is now based on the well-known and successfully tested heavy-ion collision model, HYDJET++ \cite{HYDJET_manual}. Here, we consider a modification of the statistical approach in the model, which is conceptually similar to the one proposed recently in \cite{Chernyshov_2023}, to account for the event-by-event conservation of the electric net charge of produced particles. This allowed us to reproduce the LHC data on charge balance functions measured in Pb+Pb collisions. The paper is organized as follows: in Sect.~2, we remind the reader of the main definitions and kinematic relations in the problem under consideration. The basic characteristics of the applied HYDJET++ model are outlined in Sect.~3. The simulation results and their description are presented in Sect.~4. Sect.~5 is devoted to a summary, conclusions, and discussion.

\section{Basic definitions and kinematical relations }
\label{def}

As mentioned above, alignment was observed in experiments with high-energy cosmic rays, so all kinematic, geometric, and angular relations of the detected particles in the plane of the emulsion film are referred to the laboratory coordinate system.
To characterize the azimuthal angular correlations, the Pamir Collaboration defines the alignment of spots~\cite{pamir4} in terms of the variable

\begin{equation}
\label{alig}
\lambda_{N}~=~ \frac{ \sum^{N}_{i \neq j \neq k}\cos(2 \varphi_{ijk})}
{N(N-1)(N-2)},
\end{equation}
where $\varphi_{ijk}$ is the angle between two straight lines connecting the $i$spot with the $j$th and $k$th spots. $N$ is the number of spots used in the analysis. The alignment, $\lambda_N$, quantitatively describes the deviation of points from a straight line and qualitatively characterizes asymmetry better than other possible parameters, such as eccentricity or thrust. For instance, $\lambda_4$ will be equal to 1 if all four points lie strictly on the same straight line, but it will be considerably less than 1 if these points form the four vertices of a long rectangle. The combinatorial factor $(N(N-1)(N-2))$ normalizes the parameter $\lambda_N$, as it accounts for the number of ways to choose three points from $N$. In the case of $N=3$, according to the definition in~(\ref{alig}), each angle of the triangle is counted twice, the combinatorial factor $N(N-1)(N-2)=6$ thus normalizes $\lambda_3$. For example, $\lambda_3 = -0.5$ in the case of a symmetrical configuration of three points in a plane (the equilateral triangle). It is important to note that $\lambda_N =1$ regardless of the number of points under consideration if they lie exactly along a straight line. 

The ratio of the number of events $l$ for which $\lambda_N > 0.8$ to the total number of events $L$ (in which the number of energy centers is not less than $N$) is called the degree of alignment
\begin{equation}
\label{Pn}
P_N=\frac{ l }{ L }.
\end{equation}

To estimate the maximum degree of alignment we ignore secondary interactions of particles in the atmosphere. Then, the needed azimuthal angles are calculated over the position ${\bf r}_i$ of the particles (spots) in the $(xy)$-plane in the film:
\begin{equation}
\label{position}
{\bf r}_i~=~ \frac{{\bf v}_{ri}}{v_{zi}}~h~,
\end{equation}
where  $v_{zi}$ and  ${\bf v}_{ri}$ are the longitudinal and radial components of
particle velocity respectively. $h$ is the altitude of several hundred meters to several kilometers in the Earth's atmosphere above the emulsion chamber, where the primary interaction with cosmic rays takes place.
The size of the particle observation area is about several centimeters and is related to the characteristics of the X-ray film used in the Pamir experiment. The gap for $r_i$ must satisfy the following relation:
\begin{equation}
\label{mini}
r_{\rm min}~<~r_i~<~r_{\rm max}.
\end{equation}
The condition (\ref{mini}) describes the non-coincidence of points on the film with the center $r_{\rm min}$, which is determined by different particles moving along the collision axis (mainly in the region of incident-hadron dissolution) and the location of spots within the observation film sector. The distance between particles is determined by the expression:
\begin{equation}
\label{dij}
d_{ij}~=~
 \sqrt{r^2_i~+~r^2_j~-~2r_i r_j \cos(\phi_i~-~\phi_j)}
\end{equation}
$\phi_i$ - is the the azimuthal angle, and must satisfy the condition:
\begin{equation}
\label{dijres}
d_{ij}~>~ r_{\rm min}~,
\end{equation}
which means the distinguishability of spots $i$ and $j$ from each other.
Otherwise, the particles are combined into a cluster until only particles and/or particle clusters remain with mutual distances larger than $r_{\rm min}$. The new cluster created by two particles is positioned on the emulsion film
\begin{equation}
\label{rij}
{\bf r}_{ij}=({\bf r}_i E_i+ {\bf r}_j E_j)/(E_i+E_j)
\end{equation}
and is determined as the center-of-mass coordinates of two bodies in classical mechanics where $E_i,~E_j$ - is the energy of $i$ and $j$ particles.

The alignment is calculated using expression (\ref{alig}) where, apart from the central bunch, clusters that satisfy the restrictions (\ref{mini}) and (\ref{dijres}) are considered. Among these, $2,\dots,7$ most energetic clusters/particles are selected, i.e., $N-1 = 2,\dots,7$.

\section{HYDJET++ model}
\label{HJplpl}

To the best of our knowledge, no specific searches for alignment phenomena in collider experiments have been performed so far, and, therefore, no manifestation of alignment has been found there yet. In the available data from cosmic ray experiments, there is a large measurement error. In our recent papers ~\cite{Lokhtin:2023tze,LokhtinYF24}, we attempted to understand how the selection procedure for most energetic particles, together with the energy threshold, influences the pattern of alignment. These studies are based on geometrical and kinematical considerations, not influenced by specific dynamics. Now we would like to extend this promising investigation to a more realistic model of hadron interactions. For this purpose, we use the popular and well-known HYDJET++ model, which successfully describes a large number of physical observables measured in heavy-ion collisions during RHIC and LHC operations. The details of this model can be found in the HYDJET++ manual~\cite{HYDJET_manual}.

The event generator includes two independent components: the soft, hydro-type state, and the hard state resulting from in-medium multi-parton fragmentation.
The soft component represents the ``thermal" hadronic state generated on the chemical and thermal freeze-out hypersurfaces, as prescribed by the parametrization of relativistic hydrodynamics with preset freeze-out conditions (using the adapted event generator FAST MC~\cite{Amelin:2006qe,Amelin:2007ic}). Particle multiplicities are calculated within the effective thermal volume using a statistical model approach.

The effective volume accounts for the collective velocity profile and the shape of the hypersurface, canceling out in all particle number ratios. Therefore, these ratios do not depend on the freeze-out details, provided the local thermodynamic parameters are spatially uniform. The concept of the effective volume is used to determine the hadronic composition at both chemical and thermal freeze-outs. The number of particles in an event is calculated according to a Poisson distribution around its mean value, which is assumed to be proportional to the number of participating nucleons for a given impact parameter in an A+A collision. To simulate the elliptic and triangular flow effects, a hydro-inspired parametrization for the momentum and spatial anisotropy of a soft hadron emission source is implemented~\cite{HYDJET_manual,Wiedemann:1997cr}.

In the hard sector, the model propagates hard partons through the expanding quark-gluon plasma and accounts for both the collisional loss and gluon radiation due to parton rescattering. This is based on the PYQUEN partonic energy loss model~\cite{Lokhtin:2005px}. The number of jets is generated according to a binomial distribution, with their mean number in an A+A event calculated as the product of the number of binary nucleon-nucleon (NN) sub-collisions at a given impact parameter and the integral cross-section of the hard process in NN collisions with a minimum transverse momentum transfer, $p_T^{\rm min}$. The latter is an input parameter of the model. In the HYDJET++ framework, partons produced in (semi)hard processes with momentum transfer below $p_T^{\rm min}$ are considered ``thermalized," and their hadronization products are included in the soft component of the event ``automatically".

It should be noted that, when applying this model, a ``natural" threshold for collision energy exists for the appearance of the alignment phenomenon, as the concept of quark-gluon plasma is valid only above a certain minimal collision energy.

\section{Simulation of alignment in HYDJET++ model }
\label{Modeling}

As noted in the previous section, the model calculations, such of transverse momentum spectra, pseudorapidity and centrality dependence of inclusive charged particle multiplicity, and $\pi^\pm \pi^\pm$ correlation radii in central Pb+Pb collisions~\cite{Lokhtin:2012re}, centrality and momentum dependence of second and higher-order harmonic coefficients~\cite{Bravina:2013xla}, flow fluctuations~\cite{Bravina:2015sda}, jet quenching effects~\cite{jet1,jet2}, angular dihadron correlations~\cite{Eyyubova:2014dha}, and azimuthal oscillations of femtoscopic radii~\cite{Bravina:2017rkp}, are in fair agreement with the experimental data. Therefore, to be specific, we consider Pb+Pb collisions in the centrality class $c=0-5\%$ at the center-of-mass energy of 5.02 TeV per a nucleon pair, with the free parameters of the model tuned and fixed earlier. Of course, the Pb+Pb system is much larger than the one created in interactions of cosmic nuclei in the atmosphere, such as Fe+O reactions, but it is available for study in current collider (LHC) experiments.

We believe that the alignment phenomenon can be induced by nucleus-nucleus interactions at high energy, keeping the following points in mind. First of all, the collision energy at the LHC is higher than the threshold for interactions in the Pamir experiment, where this effect emerges. Second, the creation of quark-gluon plasma is possible after a certain energy threshold, primarily in heavy-ion collisions. However, a large atomic number is not a necessary condition to observe quark-gluon plasma. Moreover, this dense matter can also be created in smaller systems, even in proton-proton collisions with high multiplicity \cite{qgpsmall, Pasechnik:2016wkt, PHENIX:2018lia}. In a sense, the consideration of Pb+Pb collisions is not crucial for our findings. In principle, one can consider collisions of lighter nuclei, but at this stage, we prefer to test our approach with Pb+Pb collisions, as they have already been experimentally investigated and are well described by the HYDJET++ model, without the need for special parameter tuning.

\begin{figure}[!h]
\centering
\includegraphics[scale=1.0]{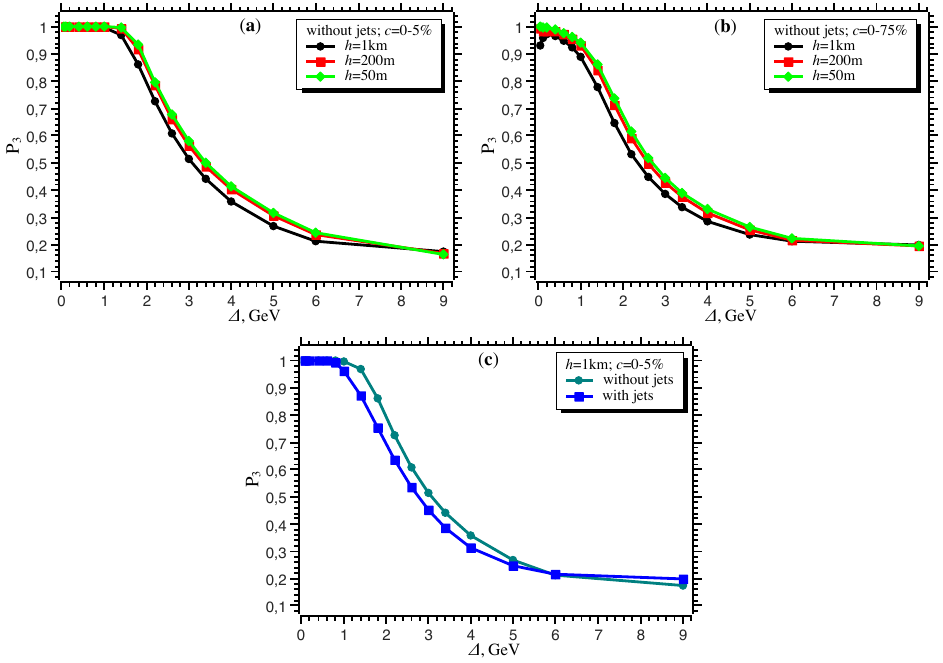}
\caption{The degree of alignment $P_3$ for three particles as a function of the disbalance $\Delta$ in agreement with the definition (\ref{Pnd}). (\textbf{a}) -- the results for only soft particles (``without jets'') at different values of the height $h$ in the centrality class collisions $c=0-5\%$; (\textbf{b}) -- the same as (\textbf{a}), but in the centrality class collisions $c=0-75\%$; (\textbf{c}) -- comparison of the results for the soft particles and the case with the jet mechanism included  (``with jets'') at the height $h=1$km.}
\label{figp3}
\end{figure}

Since the alignment was observed in the emulsion film which the laboratory frame is bound, it is convenient to write the 4-momentum of each particle $i$ under study as follows
\begin{eqnarray}
\label{momentum}
& & [\sqrt{p^2_{Ti}+m^2_i}~\cosh \eta_i,~~~ p_{Ti}\cos \phi_i,~~~ p_{Ti}\sin \phi_i, ~~~ \sqrt{p^2_{Ti}+m^2_i}~\sinh \eta_i],
\end{eqnarray}
where $p_{Ti}=\sqrt{p_{x_i}^2+p_{y_i}^2}$ is the transverse momentum (with respect to the collision axis $z$), $\eta_i$, $\phi_i$ are the pseudorapidity and the azimuthal angle in the center-of-mass system.
At the considered interaction energies, particles are ultrarelativistic, and, as is known, in this case, the values of rapidity and pseudorapidity are practically the same. For this reason, in what follows, by rapidity, we understand pseudorapidity.
An important element of our simulation is the transformation from the center-of-mass system to the laboratory frame, which is realized by the rapidity shift: $\zeta_i=\eta_0+\eta_i$ where $\eta_0$ and $\zeta_i$ are the rapidities of the center-of-mass system and the particle $i$ respectively in the laboratory reference frame. The positions $r_i$ of particles with mass $m_i$ are given as
\begin{equation}
\label{posshift}
{\bf r}_i=\frac{{\bf p}_{T_i}}{\sqrt{p_{T_i}^2 +m_i^2 } \sinh(\eta_0+\eta_i) }~h.
\end{equation}

Following the considerations in Section 2, we set the parameters $r_{\rm min}=r_{\rm res}=1$ mm, $r_{\rm max}=15$ mm, $h=1$ km, which are close to the conditions of emulsion experiments, with the additional restriction on the energy threshold of particle registration in the emulsion, $E_i > E^{\rm thr} = 4$ TeV. Our calculations are practically insensitive to this threshold within a wide interval of its variation. The results of the simulation for the degree of alignment $P_N$ in the framework of the HYDJET++ model are the following:
\begin{equation}
\label{pn}
P_3 \approx 0.2,~P_4 \approx 0.04,~P_5 \approx 0.008 ~ \rm at~ \lambda_{N} > 0.8
\end{equation}
for $N=3,4,5$ particles respectively, and are close to those obtained earlier~\cite{Lokhtin:2023tze,LokhtinYF24} for chaotically located spots in the X-ray film. This is surprising only at first glance. In fact, the azimuthal angle of particle distribution in the most central collisions is practically isotropic in the HYDJET++ generator, as is the angular distribution of points in the procedure applied in~\cite{Lokhtin:2023tze,LokhtinYF24}. The existing angular model correlations (for instance, with the reaction plane) are not sufficient to reproduce the correlations needed for the appearance of the alignment phenomenon without a special selection procedure.

Here, it is worth noting that the majority of soft particles are generated independently within the statistical model approach. In this approach, the total momentum and energy, together with the particle number, vary (fluctuate) from event to event. Only their mean values are meaningful, meaning that the transverse momentum of all particles is equal to zero not in each event, but on average. To solve this long-standing conceptual problem, the possible influence of transverse momentum conservation in every event is taken into account in the form of missing transverse momentum:
\begin{equation}
\label{a}
|{\bf p}_{T_1} + {\bf p}_{T_2} + ... + {\bf p}_{T_{(N-1)}}| < \Delta,
\end{equation}
where ${\bf p}_{T_i}$ is the transverse momentum of $i$ particle. The smaller the value of $\Delta$, the better the transverse momentum of the selected highest-energy particles should be balanced, and vice versa: the larger $\Delta$, the greater the disbalance of the transverse momentum is permissible for the selected particles. In terms of expression (\ref{Pn}), the alignment degree $P_N$ is determined for any value of $\Delta$ as
\begin{equation}
\label{Pnd}
P_{N}(\Delta)=\frac{l^{[\Delta]}}{L^{[\Delta]}},
\end{equation}
where $l^{[\Delta]}$ is the number of events with $\lambda_N > 0.8$ and $L^{[\Delta]}$ is the number of events satisfying the condition (\ref{a}). Then the modeling results can be represented as a function $P_N(\Delta)$ for the $N=3,4,5$ highest-energy particles in every event. The total number of generated events is $L^{[\rm tot]}=10^5$.

Let us take a closer look at the case $N=3$ and demonstrate the effect of disbalance $\Delta$ on the degree of alignment. For two selected most energetic particles (or clusters) and one particle in the origin the value of its total transverse momentum $\Delta_{2}$ is equal to
\begin{eqnarray}
\label{2pt}
\Delta_2= \sqrt{p^2_{T1}+p^2_{T2}+2p_{T1}p_{T2}\cos{(\phi_1-\phi_2)}}.
\end{eqnarray}
In the case of the collinear ($\phi_1-\phi_2\simeq 0$) and back-to-back ($|\phi_1-\phi_2|\simeq \pi$) configurations, one has the ``line up  events'' with a large value of the parameter $\lambda_3$. The share of such configurations among all possible ones is just $\sim 20\%$ and can be directly  estimated if the azimuthal angular distribution of particles is known.
\begin{figure}
\centering
\includegraphics[scale=1.0]{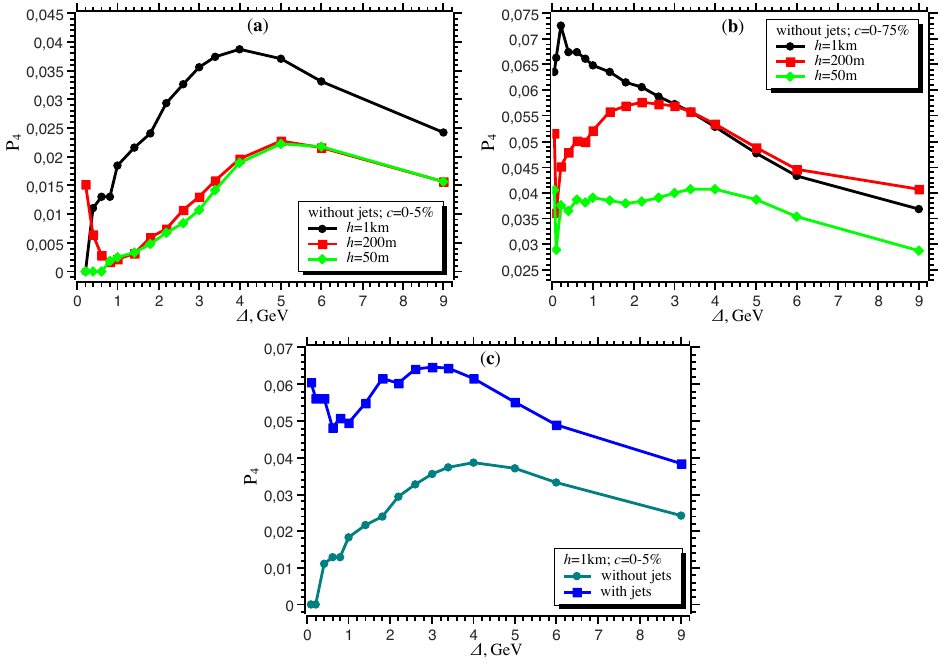}
\caption{The degree of alignment $P_4$ for three particles as a function of the disbalance $\Delta$ in agreement with the definition (\ref{Pnd}). (\textbf{a}) -- the results for only soft particles (``without jets'') at different values of the height $h$ in the centrality class collisions $c=0-5\%$; (\textbf{b}) -- the same as (\textbf{a}), but in the centrality class collisions $c=0-75\%$; (\textbf{c}) -- comparison of the results for the soft particles and the case with the jet mechanism included  (``with jets'') at the height $h=1$km.}
\label{figp4}
\end{figure}
\begin{figure}
\centering
\includegraphics[scale=1.0]{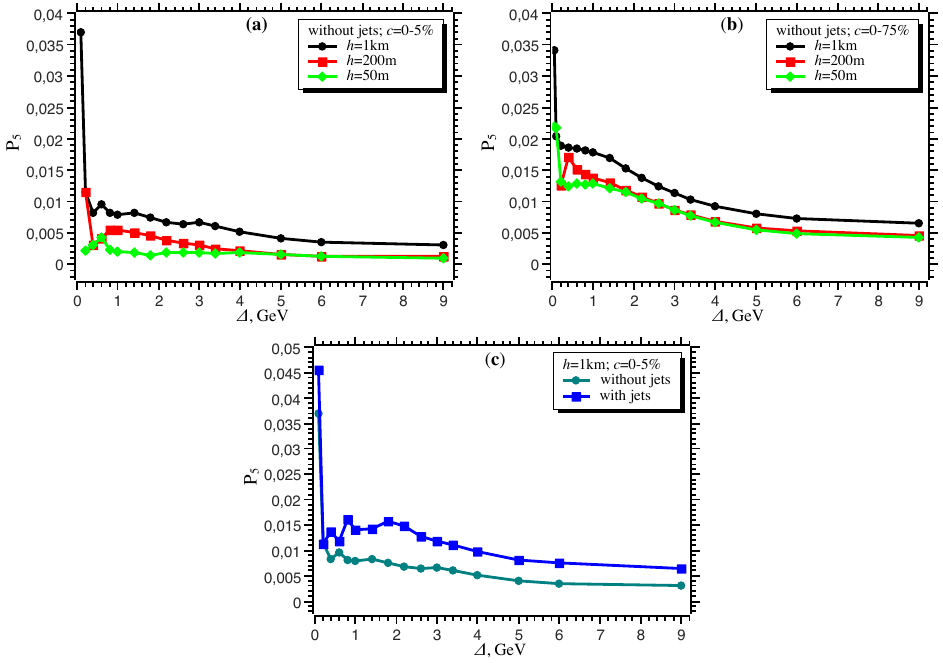}
\caption{The degree of alignment $P_5$ for three particles as a function of the disbalance $\Delta$ in agreement with the definition (\ref{Pnd}). (\textbf{a}) -- the results for only soft particles (``without jets'') at different values of the height $h$ in the centrality class collisions $c=0-5\%$; (\textbf{b}) -- the same as (\textbf{a}), but in the centrality class collisions $c=0-75\%$; (\textbf{c}) -- comparison of the results for the soft particles and the case with the jet mechanism included  (``with jets'') at the height $h=1$km.}
\label{figp5}
\end{figure}
With decreasing of $\Delta$, the condition (\ref{a}) ($\Delta_2 < \Delta$) strengthens the fraction of back-to-back configurations, since they have the minimal value of $\Delta_2 \simeq |{\bf p}_{T_1}-{\bf p}_{T_2}|$ among their possible values. The condition (\ref{mini}) about the noncoincidence of the points in the film with the center also implies that there is some minimal transverse momentum $p_{\rm min}$ of the selected highest-energy particles, according to the kinematic relation (\ref{position}). This minimal transverse momentum depends on the size ($r_{\rm min}$) of the observation region, the height ($h$) of the primary interaction, and the collision energy (via $\eta_0$). In fact, it characterizes the scale of the missing transverse momentum and provides a measure of transverse momentum resolution. Thus, the restriction on the transverse momentum from below means that at $\Delta < \sqrt{2} p_{T {\rm min}}$, only back-to-back configurations survive, and we practically achieve a $100\%$ degree of alignment. 

The peculiarity in the behavior of $P_3$ as a function of transverse momentum disbalance $\Delta$ is shown in Fig. \ref{figp3}. According to the kinematic relation (\ref{position}), the region of high degree of alignment becomes wider with decreasing height, as is also seen in Figs. \ref{figp3}a--b, because in this case the minimal transverse momentum becomes larger as well. However, this widening is not so significant since the inverse proportionality between the transverse momentum and the height in the numerator of (\ref{position}) is partially compensated by the energy decrease in the denominator.

The results of our modeling with the restriction (\ref{a}) are shown in Fig. \ref{figp4} and Fig. \ref{figp5} for four and five ($N=4$ and $N=5$) selected highest-energy particles, respectively. The degree of alignment increases noticeably by a factor of 2 for $P_4$ and by a factor of 5 for $P_5$ in comparison with their values (\ref{pn}) without the rule (\ref{a}). The effect of an odd number of particles is also present for $P_4$ and was observed in our previous studies~\cite{Lokhtin:2023tze,LokhtinYF24} for an odd number of randomly generated points. The reason is similar: the disbalance $\Delta$ for three selected particles is of an order of $p_{T {\rm min}}$ or larger if their location is close to the same line passing through the origin (center). A small value of $\Delta$ does not allow such configurations to be generated, and the degree of alignment $P_4$ is close to zero.

To disentangle the influence of different mechanisms of particle production, we first considered only the soft part of the HYDJET++ model (``without jets'' in Figs. \ref{figp3}--\ref{figp5}), which is an adapted version of the event generator FASTMC and yields the bulk of particles with transverse momentum up to $\sim 5$ GeV. In the interval of transverse momentum larger than $\sim 5$ GeV, the influence of the hard part of the considered model can become noticeable.

Figures \ref{figp3}c--\ref{figp5}c compare the results with the included jet mechanism (``with jets'') and with only soft particles. The form of the curves is practically unchanged, but a significant increase in the alignment degree occurs for four and five particles analyzed. Moreover, the effect of an odd particle number is less pronounced for $P_4$, since a small disbalance $\Delta$ is achievable for back-to-back configurations with a large enough value of $\lambda$, because the transverse momentum of one jet particle can be easily compensated by the total transverse momentum of two soft particles with relatively smaller characteristic momentum.

To investigate the possible influence of elliptic and triangular flow effects (included in the event generating mode) on the alignment degree $P_N$, we also consider the so-called ``minimum bias", specifically centrality class collisions $c=0-75\%$ (see Figs. \ref{figp3}b--\ref{figp5}b). In addition, at $c=0-75\%$, the modeling parameters are as close as possible to the conditions of experiments with cosmic rays. It is challenging to evaluate the impact of anisotropic flow on alignment with the jet mechanism activated, as the contribution of jets to the overall multiplicity significantly depends on centrality, which introduces additional uncertainty into the simulation results. Therefore, it would be reasonable to compare the results for $P_N$ for soft particles only, i.e., without jets. The results for $P_3$ and $P_5$ remain practically unchanged, but for $P_4$, we observe a less pronounced effect of the odd number of considered particles, as is shown in Fig. \ref{figp4}b. We associate this with the manifestation of transverse momentum equality enhanced by anisotropic flow of one particle with the total transverse momentum of two ``softer" particles, each having a relatively smaller characteristic momentum. Therefore, high alignment values can only be realized with relatively low levels of missing momentum $\Delta$.

\section{Discussion and Conclusions}
\label{ConclDisc}

Presumably, the alignment can manifest itself in both high-energy collisions of relatively light nuclei (in cosmic ray experiments) and heavy ions (in collider experiments), together with collective effects such as the quark-gluon plasma formation and local conservation of the transverse momentum of detected particles and their clusters. We have analyzed the influence of the selection procedure for the most energetic particles on the appearance of the alignment phenomenon within the framework of the HYDJET++ model. Event by event, the transverse momentum conservation has been taken into account in the form of missing transverse momentum (\ref{a}), since the majority of particles are independently generated in the statistical model approach. The results in Figs. \ref{figp3}--\ref{figp5} show that a high degree of alignment appears at reasonable values (up to $\sim3$ GeV) of the transverse momentum disbalance of the selected most energetic particles. In addition, the HYDJET++ model assumes statistical particle production due to the hadronization of the quark-gluon plasma, which automatically implies a collision energy threshold. This threshold is also characteristic of alignment observation.

\begin{figure}[!h]
\centering
\includegraphics[scale=0.45]{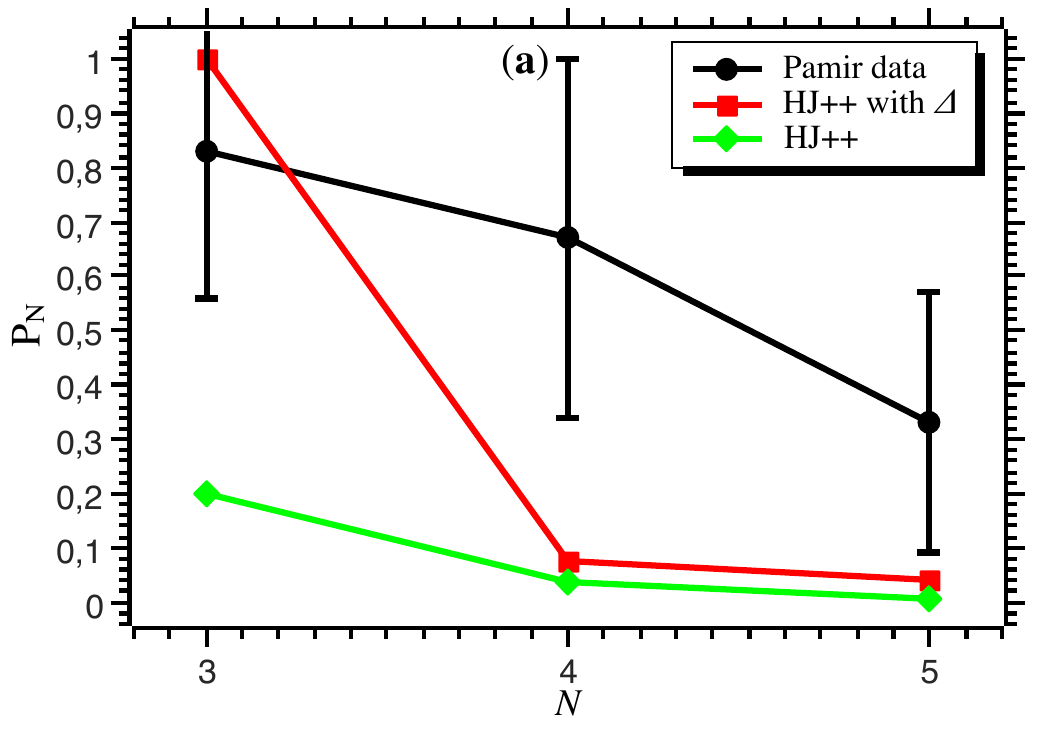}
\includegraphics[scale=0.45]{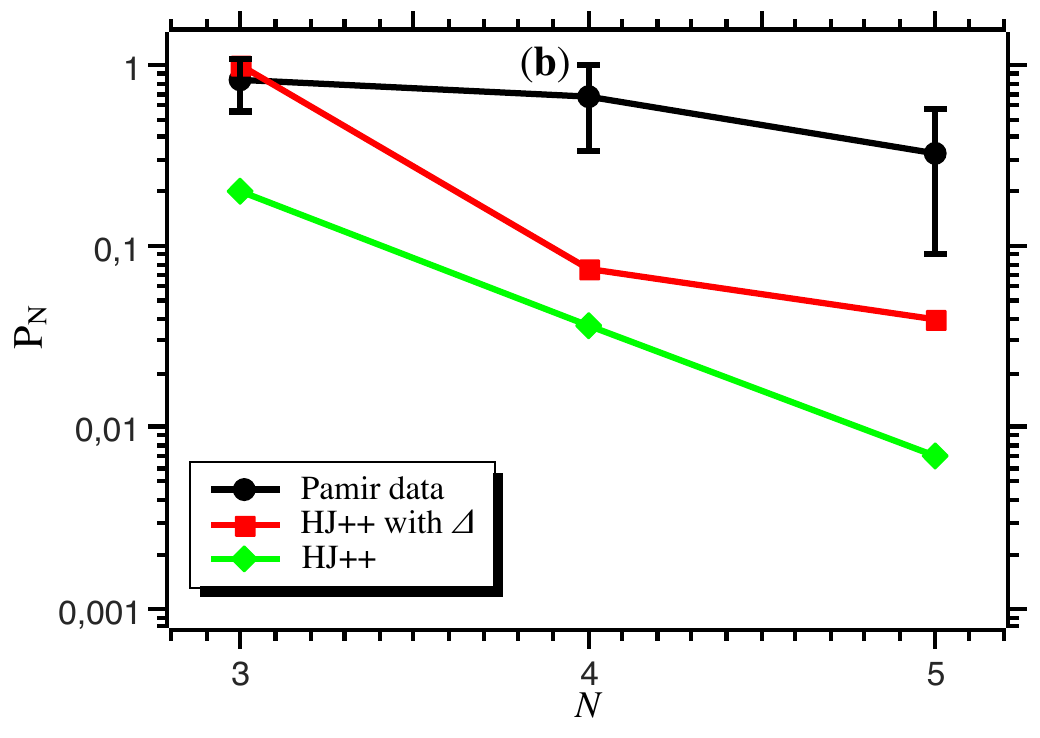}
\caption{Comparison of the simulation results of the alignment degree $P_N$ for three, four, five particles with the data of the Pamir collaboration. (\textbf{a}) -- linear scale, (\textbf{b}) -- logarithmic scale. ``HJ++ with $\Delta$'' -- the modeling with taking account the local transverse momentum conservation with disbalance $\Delta$ (\ref{a}); ``HJ++'' -- the modeling without ${\bf p}_T$ conservation in every event. The values of the disbalance $\Delta$, which correspond to the our ``best'' alignment degree, are in the range of $0 \div 1$ GeV.}
\label{figcomp}
\centering
\end{figure}

\begin{table}
\caption{The estimation of the fractions of the selected events $L^{[\Delta]}$ out of the total interaction rate $L^{[\rm tot]}=10^5$ is provided for each $\Delta$ and for $N=3,4,5$ highest-energy particles in every event. The ratios for only soft particles are shown (``without jets'') at a height of $h=1$ km in centrality class collisions $c=0-5\%$.}
\label{tab1}%
\begin{ruledtabular}
\begin{tabular}{ c c c c c c c c c c c c c c c c c c c c c c }
& & & & \multicolumn{3}{ c }{$N=3$} & & & \multicolumn{3}{ c }{$N=4$} & & & \multicolumn{3}{ c }{$N=5$} & & \\
\cmidrule{5-7} \cmidrule{10-12} \cmidrule{15-17}  
& $\Delta$, GeV & & & &$L^{[\Delta]} / L^{[\rm tot]}$ & & & & & $L^{[\Delta]} / L^{[\rm tot]}$ & & & & & $L^{[\Delta]} / L^{[\rm tot]}$ & &\\
\midrule
& $0.05$  & & & & $2*10^{-4}$ & & & & & $7*10^{-5}$ & & & & & $4*10^{-5}$ \\
& $0.1$ & & & & $4*10^{-4}$ & & & & & $3*10^{-4}$ & & & & & $2*10^{-4}$  \\
& $0.4$ & & & & $7*10^{-3}$ & & & & & $4*10^{-3}$ & & & & & $4*10^{-3}$  \\
& $0.8$ & & & & $3*10^{-2}$ & & & & & $15*10^{-3}$ & & & & & $13*10^{-3}$  \\
& $1.0$ & & & & $4*10^{-2}$ & & & & & $2*10^{-2}$ & & & & & $2*10^{-2}$  \\
& $3.0$ & & & & $0.3$ & & & & & $0.2$ & & & & & $0.2$  \\
& $5.0$ & & & & $0.6$ & & & & & $0.5$ & & & & & $0.4$  \\
& $9.0$ & & & & $0.9$ & & & & & $0.9$ & & & & & $0.8$  \\
\end{tabular}
\end{ruledtabular}
\end{table}

A comparison of the simulation results with the Pamir collaboration data is shown in Fig. \ref{figcomp}. It should be noted that, since the Pamir data were obtained from experiments with cosmic rays, while our results are related to heavy-ion collisions, only a qualitative comparison can be made. Figure \ref{figcomp} shows that the alignment degree of three particles ($N=3$) can be fully described within the HYDJET++ model, taking into account the local transverse momentum conservation with the disbalance $\Delta$, whereas for four ($N=4$) and five ($N=5$) particles, we do not reproduce the experimental data \cite{man}, even when measurement error is taken into account:
\begin{gather}
P_3^{\rm exp}=0.83 \pm 0.27; ~ P_4^{\rm exp}=0.67 \pm 0.33; ~ P_5^{\rm exp}=0.33 \pm 0.23, \nonumber
\end{gather}
while our ``best'' values are
\begin{equation}
P_3=1; ~ P_4 \approx 0.07; ~ P_5 \approx 0.04 , \nonumber
\end{equation}
which are achieved at the disbalance in the interval
\begin{equation}
\Delta=0 \div 1 ~\text{GeV}, \nonumber
\end{equation}
corresponding to the transverse momentum conservation with high enough accuracy. Nevertheless the alignment degrees $P_4$ and $P_5$  are in the limits of the two standard deviations from the experimentally measured values as a minimum, in addition, the values $P_4$ and $P_5$ increase by a factor of about 2 and 5, respectively, compared to result (\ref{pn}) due to $p_{T}$-conservation in each event.

The introduction of $\Delta$ implicitly means the selection of a subset of events $L^{[\Delta]}$ modeled by the HYDJET++ program from the total number of interactions $L^{[\rm tot]}$. Since the HYDJET++ model does not include explicit event-by-event conservation of transverse momentum, but only does so on average, i.e., in some rapidity interval, the introduction of the parameter $\Delta$ allows us to effectively take into account such $p_{T}$-conservation with an accuracy of $\sim \Delta$ at the level of selected/rejected events, and this leads to some loss in the total number of generated events. In the case of two selected most energetic particles ($N=3$ taking into account central cluster), for instance, the proportion of selected events $L^{[\Delta]}$ relative to the total generated number $L^{[\rm tot]}$ can be roughly estimated (if $\Delta$ goes to zero) as proportional to $ \Delta \big /<p_{T}> $, where $<p_{T}>$ is the mean particle transverse momentum, which is determined for the soft component of the model by the thermal freeze-out temperature $ T \sim 100 $ MeV and the collective radial velocity. At values of $\Delta$ larger than $<p_{T}>$, the number of selected events rapidly approaches the number of generated ones. Thus, the fraction of rejected events is significant only in the region $\Delta$ less than $<p_{T}>$. This fact can partly justify our solution to the problem of transverse momentum conservation in the statistical model approach. The ratio of the number of events $L^{[\Delta]}$ in each $\Delta$ value to the total number of generated events $L^{[\rm tot]}$ is shown in Table \ref{tab1}, which demonstrates that with decreasing parameter $\Delta$ the number of selected events $L^{[\Delta]}$ also decreases, and vice versa: as $\Delta$ increases, the number of $L^{[\Delta]}$ tends to $L^{[\rm tot]}$, as expected.

Thus, our modeling clearly demonstrates that the conservation law of transverse momentum, via minimization of the disbalance $\Delta$ of the total transverse momentum of the highest-energy particles, allows one to automatically select more aligned configurations. The application of this procedure to the HYDJET++ model can be partially justified by the fact that the majority of soft particles are generated independently, as mentioned above, and that transverse momentum is conserved only on average, but within any rapidity interval. However, one should note that the degree of alignment $P_N$ in our modeling is not large enough to match the central values of experimental measurements for $N \geq 4$, leaving some room for other explanations and further investigations.

\acknowledgments

We are grateful to A.S.~Chernyshov, A.I.~Demianov for fruitful discussions and S.N. Nedelko for our communication. This work is supported by the Russian Science Foundation, grant 24-22-00011.


\end{document}